\documentclass[a4paper,fleqn,usenatbib]{mnras}

\usepackage{newtxtext,newtxmath}
\usepackage[T1]{fontenc}
\usepackage{ae,aecompl}

\usepackage{graphicx}	% Including figure files
\usepackage{amsmath}	% Advanced maths commands
\usepackage{amssymb}	% Extra maths symbols

\defcitealias{Marmol09}{MQ09}
\defcitealias{Silva08}{S08}
\title[The interpretation of NIR indices: NaI2.21]
  {The puzzling interpretation of NIR indices: The case of NaI2.21}
\author[B. R\"{o}ck et al.]
  {B. R\"{o}ck$^{1,2}$,
   A. Vazdekis$^{1,2}$,
   F. La Barbera$^3$,
   R.F. Peletier$^4$,
   J.H. Knapen$^{1,2}$,  
\newauthor
 C. Allende-Prieto$^{1,2}$ and D. S. Aguado$^{1,2}$  
   \\
  $^1$Instituto de Astrof\'{i}sica de Canarias, V\'{i}a L\'{a}ctea s/n, E-38205 La Laguna, Tenerife, Spain\\
  $^2$Departamento de Astrof\'{i}sica, Universidad de La Laguna, E-38205 La Laguna, Tenerife, Spain\\
  $^3$INAF-Osservatorio Astronomico di Capodimonte, sal. Moiariello 16, Napoli I-80131, Italy\\
  $^4$Kapteyn Astronomical Institute, University of Groningen, Postbus 800, 9700 AV, Groningen, The Netherlands}

\date{Released 2016 Xxxxx XX}

\pagerange{\pageref{firstpage}--\pageref{lastpage}} \pubyear{2016}

\def\LaTeX{L\kern-.36em\raise.3ex\hbox{a}\kern-.15em
    T\kern-.1667em\lower.7ex\hbox{E}\kern-.125emX}

\begin{document}

\label{firstpage}

\maketitle

\begin{abstract}

We present a detailed study of the Na I line strength index centered in the $K$-band at ${\rm 22100 \, \AA}$ (NaI2.21 hereafter) relying on different samples of early-type galaxies. Consistent with previous studies, we find that the observed line strength indices cannot be fit by state-of-art scaled-solar stellar population models, even using our newly developed models in the NIR. The models clearly underestimate the large NaI2.21 values measured for most early-type galaxies. 
%We show that neither variations in age and metallicity nor in the fraction of asymptotic giant branch stars contributing to our models can explain conclusively the indices measured for the galaxies. Only for cool dwarf stars we obtain equally large NaI2.21 values as for the galaxies. 
However, we develop a Na-enhanced version of our newly developed models in the NIR, which - together with the effect of a bottom-heavy initial mass function - yield NaI2.21 indices in the range of the observations. Therefore, we suggest a scenario in which the combined effect of [Na/Fe] enhancement and a bottom-heavy initial mass function are mainly responsible for the large NaI2.21 indices observed for most early-type galaxies. To a smaller extent, also [C/Fe] enhancement might contribute to the large observed NaI2.21 values.

\end{abstract}

% On the other hand, absorption due to interstellar gas and dust does not play an important role in the $K$-band. 

%Both \citet{Spiniello12} and \citet{Jeong13} reach a very similar conclusion in the case of the Na D feature in the optical spectral range. They explain the large observed EWs by the combined effect of [Na/Fe], $\alpha$ and [Fe/H] enhancement and a more bottom-heavy IMF. Furthermore, \citet{Smith15b} argued that the large Na1.14 indices of ETGs are mainly driven by a more bottom-heavy IMF and [Na/Fe] enhancement. 


\section{Introduction}\label{introduction}

Studying the absorption features in the spectra of unresolved stellar populations is crucial to assess their stellar content and chemical composition. We are however still lacking a comprehensive picture based on the line strength indices from many different elements, which would allow us to put some closer constraints on the way galaxies formed and evolved. 

Among all these absorption lines, the origin and behaviour of the four strong sodium features, which are found in the optical and NIR spectral region, is particularly puzzling. The four Na absorption features in the optical and NIR wavelength range encompass the {4 doublets situated at 5890 and ${\rm 5896 \, \AA}$ (NaD hereafter), at 8183 and ${\rm 8195 \, \AA}$ (NaI 0.82 hererafter), at 11381 and ${\rm 11404 \, \AA}$ (NaI1.14 hereafter), as well as at 22062 and ${\rm 22090 \, \AA}$ (NaI2.21 hereafter), respectively. Particularly the latter two Na features have been poorly studied in the literature so far.

As pointed out by various authors \citep{OConnell76, Peterson76, Worthey98, vanDokkum10, Conroy12, Spiniello12, Jeong13, Smith15}, sodium, which is believed to be produced mainly in SNII, causes very strong absorption features in galaxies superseding by far the values measured for most stars and stellar population models. Since Na is supposedly also produced in asymptotic giant branch (AGB) stars \citep{Mowlavi99}, analyzing Na features in galaxies might also lead to a better understanding of the importance of this poorly constrained stellar evolutionary phase for the total stellar content of galaxies.

For the first time, \citet{OConnell76} and \citet{Peterson76} reported that the index values of NaD, which they measured for early-type galaxies, were much larger than what was expected from Ca and Fe indices. Later, \citet{Worthey98} finds enhanced Na indices for his sample of large elliptical galaxies and spheroids. While this author attributes this behaviour to an overabundance of [Na/Fe] compared to scaled-solar values, the NaD index is also found to be sensitive to age, metallicity, the interstellar medium (ISM) and the initial mass function (IMF) \citep[see, e.g.,][]{Chen10, Conroy12, Spiniello12, Jeong13}. Recently, also \citet{Spiniello15} concluded that the NaD index seems to be driven primarily by the [Na/Fe] abundance, and additionally to a certain extent by interstellar absorption.

On a stellar level, the NaI 0.82  doublet has long been known to be much more prominent in cool, low-mass M dwarfs than in giants \citep[e.g.,][]{Spinrad71, Carter86}. Hence, this index constitutes a great opportunity to constrain the low-mass end of the stellar IMF in unresolved stellar populations. Also \citet{vanDokkum10, LaBarbera13, Ferreras13} ascribe the large values of the NaI 0.82 doublet which they measure for massive early-type galaxies to a bottom-heavy IMF, i.e. to an excess of low- to high-mass stars in these galaxies with respect to the distribution in the Milky Way. Likewise, \citet{Spiniello15} find the NaI 0.82 doublet to be more sensitive to the underlying IMF than the NaD.

\citet{Smith15} discusses the Na I doublet at ${\rm 1.14 \, \mu m}$ in early-type galaxies. They are able to account for the strength of this feature with theoretical stellar population models, when allowing high [Na/H] values as well as a very bottom-heavy IMF. However, \citet{Smith15} obtain a less extreme IMF when applying a full-spectral fitting approach, with a couple of galaxies also showing strong-lensing masses that are inconsistent with a bottom-heavy IMF.

In early-type galaxies, similarly large line strength indices as for the other Na-features are observed for the pronounced NaI2.21 sodium line located in the $K$-band \citep{Silva08, Cesetti09, Marmol09}. The same strong NaI2.21 features as found in galaxy spectra were previously only measured by \citet{Foerster00} for her sample of supergiants and later confirmed by \citet{Rayner09} for supergiants and also for low mass, low-temperature dwarfs. However, for hotter dwarfs and giants, the latter obtain significantly smaller indices of the NaI2.21 than for galaxies and cool stars. These smaller indices of hotter dwarfs and giants are comparable with those found by \citet{Ramirez97} for a sample of disk giants as well as by \citet{Silva08} for K and M giants in Galactic open clusters.

Contrary to the situation for the NaD, absorption caused by the ISM does not play a role for the NaI2.21, since it is not a resonance line. Thus, studying the NaI2.21 might enable us to obtain a better understanding of what kind of drivers are mainly responsible for the observed excesses of Na in early-type galaxies. Furthermore, we have computed a new set of single stellar population (SSP) models in the NIR based on the empirical stellar IRTF library \citep{Roeck15}, for which we also developed a Na-enhanced version. In this paper, we take advantage of these new models to analyze this index for an unprecedentedly large compendium of samples of early-type galaxies which we adopted from the literature. 

The outline of the paper is the following. In Section \ref{data}, we describe briefly all samples of early-type galaxies as well as all the tools which we used to interpret our findings. These encompass our new stellar population models in the NIR, an auxiliary sample of theoretical stellar spectra and the Na-enhanced version of our models, which make use of those spectra. We discuss the behaviour of the NaI2.21 index in early-type galaxies and propose a possible explanation for it in Section \ref{results}. Finally, we sum up the most important points in Section \ref{conclusion}.

%\section{Stellar data and SSP models}

%In this Section, we briefly describe the IRTF stellar library \citep{Cushing05, Rayner09} as well as our new extended MILES SSP models in the infrared (IR) (Roeck et al., 2015, submitted to MNRAS, R15b hereafter) which are based on the stars from this library. 

\section{Models and data}\label{data}

Here, we briefly describe the different samples of theoretical and observed stars and galaxies which we used for our analysis. Moreover, we shortly summarize our newly developed SSP models in the NIR and the 180 stars of the IRTF library on which they are based \citep{Roeck16} as well as their [Na/Fe]-enhanced version, see La Barbera et al., 2016b, accepted for publication in MNRAS. To measure the various line strength indices, we convolved all spectra to a common resolution of ${\rm \sigma = 360 \, km \, s^{-1}}$, which is the upper limit of the individual velocity dispersions of the different galaxies.

\subsection{IRTF library and stellar population models}

We developed new single stellar population models covering the wavelength range between 8150 and ${\rm 50000 \, \r{A}}$ at a resolution of ${\rm \sigma = 60 \, km s^{-1}}$. 
These models use as ingredients 180 observed, mostly bright, nearby and cool, stellar spectra of the IRTF library \citep{Rayner09}. This renders them much more accurate than most existing models which are based on theoretical stellar libraries like, e.g., the models by \citet{Maraston05} and \citet{ Maraston09}. The CvD-models, on the other hand, are based on a smaller subsample of only around 90 IRTF stars compared to our models. However, due to the limited number of stars of mostly around Solar metallicity in the IRTF library, the resulting models are only reliable for metallicities between ${\rm [M/H] = -0.40}$ and ${\rm [M/H] = 0.20}$ and ages larger than 1 Gyr. We computed these new stellar population models based on both BaSTI \citep{Pietrinferni04, Pietrinferni06} and Padova \citep{Girardi00} isochrones as well as for various types of initial mass functions (IMF) \citep[see][]{Roeck15}. We additionally combined these new models in the IR with the already existing models in the optical which are based on the MILES library \citep{Vazdekis10, Vazdekis12}. The resulting models cover the whole optical and IR wavelength range between 3500 and ${\rm 50000 \, \r{A}}$ (R\"{o}ck et al., 2016). The new models are available to the public on the MILES web page (http://miles.iac.es).

\subsection{Theoretical stellar spectra}\label{indices_Na_Carlos}

For the present study, we computed theoretical stellar spectra for nine different $T_{{\rm eff}} - \log(g)$ combinations corresponding to typical stages of stellar evolution which are important in the NIR (see Table \ref{Stars_Carlos}). Hence, all of these theoretical stellar spectra have rather cool temperatures. The stellar spectra were calculated based on Kurucz model atmospheres \citep{Meszaros12} and the synthesis code ASS$\epsilon$T \citep{Koesterke08, Koesterke09} for the wavelength range between 3000 and ${\rm 30000 \, \r{A}}$, and at wavelength steps of ${\rm 0.6 \, km \, s^{-1}}$, which are constant with $\lambda$. Moreover, line absorption is included from the atomic and molecular (H$_2$, CH, C$_2$, CN, CO, NH, OH, MgH, SiH, SiO and TiO) files of Kurucz.
%but for upgraded van der Waals damping constants according to \citet{Barklem00}. 
The nine theoretical stellar spectra were calculated for Solar abundances as well as for [C/Fe]=0.15, since we find a strong correlation of the NaI2.21 indices with carbon-sensitive features of galaxies in the optical range (see Section \ref{indices_Na_drivers}). For both cases, also a version with an enhancement of ${\rm +0.2 \, dex}$ in [Na/Fe] was computed. 

\subsection{Na-enhanced stellar population models}\label{models_Na_enhanced}

In a second step, we calculated Na-enhanced theoretical stellar spectra corresponding to the parameters of the 180 stars of the IRTF library according to \citet{Meszaros12}. We obtained the model atmospheres by interpolating the ODFNEW4 ATLAS99 models of \citet{Castelli03}. Since these models are only available for stars with effective temperatures higher than 3500 K, we adopted a fixed temperature of 3500 K for all cooler stars in the IRTF library. For further details about the modelling, we refer the reader to both \citet{Allende08} and La Barbera et al., 2016b.

%These models use the reference solar composition from \citet{Grevesse98}. For stars with [Fe/H]<-1, we apply models which are $\alpha$-enhanced by 0.4 dex. The spectral synthesis was performed with the ASS$\epsilon$T code \citep{Koesterke08, Koesterke09}, with updates to atomic damping constants published by \citet{Barklem00}, including transitions of H2,CH, C2, CN, CO, NH, OH, MgH, SiH, and SiO. The equation of state included the first 92 elements in the periodic table and 338 molecules. We accounted for bound-free absorption from H, H- ,HeI and HeII, and the first two ionization stages of C, N, O, Na, Mg, Al, Si, and Ca from the Opacity Project \citep[see, e.g.,][]{Cunto93} and Fe from the Iron Project \citep{Nahar95, Bautista97}. All models are computed under the assumption of Local Thermodynamical Equilibrium (LTE). We refer the reader to \citet{Allende08} for more details.}

We computed the synthetic stellar spectra for a range of [Na/Fe], from 0 to +1.2 dex in steps of 0.3 dex, covering the spectral range from 0.35 to ${\rm 2.5 \, \mu m}$.  For a given value of [Na/Fe], we divided each Na-enhanced, theoretical, spectrum by its scaled-solar counterpart, to obtain the (multiplicative) differential response to [Na/Fe], which is applied to the empirical stellar spectrum. This procedure was applied to all the stars in the IRTF library. For the correction for the other stellar libraries (MILES, Indo-US, CaT), we applied our local interpolation algorithm to obtain the corresponding Na responses. We smoothed the theoretical stellar spectra, adjusting the effective resolution to match the instrumental configurations of the different stellar libraries. The spectral resolution is kept constant with wavelength (at ${\rm FWHM=2.5 \, \AA}$), for all libraries, except for the IRTF one, which is characterized by a constant ${\rm \sigma = 60 km s^{-1}}$.

Finally, we obtained our various sets of Na-enhanced stellar population models based on the modified stellar spectra including the differential corrections for the different [Na/Fe] overabundances. We refer the interested reader to La Barbera et al., 2016b, for details about the whole procedure.

\begin{table}
\caption{ Stellar atmospheric parameters of nine theoretical stellar spectra used as a reference in the present analysis.}
\label{Stars_Carlos}
\centering
\begin{tabular}{l c }
\\
\hline
 $T_{{\rm eff}}$ [K]& $\log(g)$ [dex]\\
\hline
        3500  &    0.5   \\
        3800  &    1.0    \\ 
        4000  &    1.5    \\
        4200  &    2.0    \\  
        4500  &    2.5    \\   
        4800  &    3.5    \\  
        5500  &    4.5    \\
        4000  &    4.7    \\
        3500  &    5.0    \\
  \hline
\end{tabular}
%\tablefoot{
%\medskip
\end{table}

\subsection{Samples of observed early-type galaxies}\label{indices_data_Silva}

We combined different samples of observed early-type galaxies drawn from the literature. These samples encompass different spectral ranges and have been observed with different instruments at very different resolutions. Therefore, they are far from homogeneous. The various samples are:

(i) six elliptical galaxies and three S0 galaxies in the Fornax cluster which were observed within an aperture of radius $ 1/8 \, r_{{\rm eff}}$ along the observed position angle using a 1 arcsec wide slit by \citetalias{Silva08} using the ISAAC NIR imaging spectrometer at the ESO Very Large Telescope (VLT). Then, the spectra were integrated along this aperture with an unweighted spatial sampling. The observed spectral range extends from 21200 to ${\rm 23700 \, \AA}$.  

(ii) twelve elliptical and S0 field galaxies which were observed by \citet[][hereafter MQ09]{Marmol09} using the ISAAC NIR imaging spectrometer with the same configuration as used by \citetalias{Silva08} for the Fornax cluster galaxies. 

(iii) 12 pre-reduced spectra of nearby ($z < 0.01$) quiescent spiral galaxies of mostly Hubble types S0 and Sa in the $H$- and $K$-band observed by \citet{Kotilainen12} using the 3.5 m ESO New Technology Telescope equipped with the 1024 x 1024 pixel SOFI camera and spectrograph \citep{Moorwood98}. \citet{Kotilainen12} extracted their spectra within an aperture of 1 arcsec diameter around the center of the galaxies, which corresponds to a region of around ${\rm 1 \, kpc}$ at the average distance of their galaxy sample. They also used a slit width of 1 arcsec, along which they summed up their spectra. From the original sample of 29 galaxies, we had to exclude the spectra for which no velocity dispersion $\sigma$ was available. %Moreover, we also discarded two further galaxies which show some peculiar features. 
We further constrained the remaining sample of galaxies by obtaining their \textit{WISE} $(W1-W2)$ and $(W3-W4)$ colours from the All WISE catalogue and removing all galaxies with $ (W1-W2)>0.1 \, {\rm mag}$ and $ (W3-W4)>1.8 \, {\rm mag}$. This approach is justified by the finding that a red \textit{WISE} $(W1-W2)$ colour is an indicator for central star formation in galaxies \citep{Shapiro10, Roeck15}. The finally selected sample of \citet{Kotilainen12} includes a number of spiral galaxies with lower, near-Solar metallicity, which are of particular interest here when comparing Solar-metallicity models with the data.

(iv) 9 elliptical and spiral galaxies observed by Cesetti et al. using the X-Shooter multi-wavelength medium resolution slit echelle spectrograph \citep{Vernet11} at the ESO-VLT. Also these spectra were extracted within an aperture compatible to the other samples of galaxies using a slit width of 1 arcsec, along which the spectra were summed up in an unweighted manner. From the ESO archive, we obtained the version of the galaxy spectra which had been preprocessed by the X-Shooter pipeline. Then, we carried out a correction for telluric features based on the libraries of telluric model spectra presented in \citet{Noll12} following a procedure similar to the one suggested by \citet{Moehler14}. From the complete sample of 14 galaxies, we finally had to exclude three galaxy spectra since they were severely hampered by spikes. These spikes are most likely residuals which were artificially introduced during the pre-procession by the X-Shooter pipeline \citep[see][]{Chen14}. For one more spectrum it was impossible to obtain a satisfactory correction for tellurics and for another spectrum no central velocity dispersion was available in the literature. That is why we ended up with a final sample of nine spectra (called Cesetti-galaxies hereafter).

(v) two massive SDSS-galaxies with ${\rm \sigma \approx 300 \, km s^{-1}}$ observed by \citet{LaBarbera16a} using also X-Shooter.\citet{LaBarbera16a} extracted these spectra within an aperture of ${\rm \pm 0.675 \, arcsec}$ around the centers of the galaxies using a 0.9 arcsec-wide slit. Also \citet{LaBarbera16a} just summed up the spectra along the slit within the given aperture. Since the effective radius of these two galaxies is ${\rm \approx 4.1 - 4.2 \, arcsec}$, the aperture translates into $ \approx 1/6 \, r_{{\rm eff}}$. \citet{LaBarbera16a} had selected these galaxies from the SPIDER survey \citep{LaBarbera10} according to their abundance ratios ${\rm [\alpha/Fe]}$. While they chose their first galaxy XSG1 to have a high abundance ratio of ${\rm [\alpha/Fe] \approx 0.4 \, dex}$, their second galaxy XSG2 shows a ratio of ${\rm [\alpha/Fe] \approx 0.25 \, dex}$ which is more typical for its velocity dispersion. They performed the main data reduction based on version 2.4.0 of the X-Shooter data reduction pipeline \citep{Modigliani10}. For the flux calibration, sky subtraction, telluric corrections and combination of the various exposures, \citet{LaBarbera16a} used their own routines. 

The measured line strength indices together with the central velocity dispersions $\sigma$, ages, metallicities and abundance ratios are shown in Tables \ref{measurements_I_Table} and \ref{measurements_II_Table} in the appendix. 

\section{Results}\label{results}

\subsection{Comparison of NaI2.21 in early-type galaxies to NaI2.21 in single stars}\label{indices_Na_stars}

%\begin{figure}[!t!]
%\begin{center}
% \resizebox{\hsize}{!}{\includegraphics{NaI_versus_Teff_logg_FeH_IRTF.eps}}
%  \caption{Illustration of the EWs of the Na2.21 index measured from the single stars of the IRTF-library at a resolution of ${\rm %\sigma = 360 \, kms^{-1}}$ as a function of their effective temperature ${\rm \textit{T}_{eff}}$ (left-hand panel), surface gravity log({\textit g})(middle panel) and metallicity [Fe/H] (right-hand panel). The different symbols denote the various luminosity types of stars. While the red plus signs, five pointed stars and circles stand for carbon stars, AGB stars and supergiants, respectively, the black triangles mark giant stars. Dwarf stars are visualized by blue squares. The filled symbols denote the theoretical stellar spectra of Solar and supersolar metallicities of [Fe/H]=0.2, respectively. Theoretical stellar spectra of the same $T_{{\rm eff}}$ and $\log(g)$ are linked by a black line. For the two sets of theoretical stellar spectra displayed here, other element abundances were not changed with respect to the Solar-scaled values.}\qquad
%   \label{NaI_SAPs}
%\end{center}
%\end{figure}

To reach a better understanding of the behaviour of the NaI2.21, we analyze this index for all different kind of stars in the IRTF library. \citet{vanDokkum10}, among others, showed that the NaI 0.82 increases significantly as a function of rising surface gravity at fixed effective temperature. For the NaD, \citet{Spiniello14} measured the values of the NaD feature for all the stars of the MILES library. They found the largest NaD values for low temperature dwarf stars (see their Fig. 2). Based on the stars of the IRTF library, \citet{Rayner09} report a strong increase of the NaI2.21 index as a function of decreasing effective temperature for both dwarfs, giants and supergiants, respectively (see their Fig. 35, bottom-right panel). These larger index values of NaI2.21 for cooler stars compared to their hotter companions have been extensively described in the literature \citep{Kleinmann86, Ali95, Ramirez97, Foerster00}. Moreover, \citet{Rayner09} also observe the stronger NaI2.21 features of supergiants with respect to giants of the same temperature which have been reported before by \citet{Foerster00}. However, \citet{Rayner09} found the largest indices of NaI2.21 for a subsample of about half of the low-temperature dwarfs, which exhibit much larger values than the bulk of the AGB stars and the rest of the cool dwarfs.

Furthermore, \citetalias{Silva08} and by \citetalias{Marmol09} reported the generally large indices for the NaI2.21 of ETGs compared to, e.g., those of K and M giants observed in Galactic open clusters for their samples of ETGs. Also, \citet{Cesetti09} find this excess in the index values of the NaI2.21 in their sample of 19 observed early-type galaxies.

%A similar increase is obtained for the Na2.21 from the models by \citet[][hereafter CvD models]{Conroy12} when moving from models based on a Salpeter-like IMF to a more bottom-heavy one. 

%\begin{figure}
%\begin{center}
% \resizebox{\hsize}{!}{\includegraphics{NaI_K_versus_age_sigma_letter_BB.eps}}
%  \caption{Left-hand panel: Behavior of the Na2.21 index as a function of age. The solid lines display the predictions from our (BaSTI-based) models for three different metallicities and an underlying Kroupa-like IMF, whereas the dashed black line illustrates the respective EWs measured from a model of Solar metallicity and bottom-heavy IMF of slope ${\rm \Gamma_b = 2.8}$ (see also legend). Overplotted are the EWs of Na2.21 for the early-type galaxies of \citetalias{Marmol09} (black dots), of \citetalias{Silva08} (red triangles) and for the X-Shooter galaxies of Cesetti (green squares). Right-hand panel: Behavior of the Na2.21 index as a function of central velocity dispersion $\sigma$ for the same three sample of galaxies.}\qquad
%\label{NaI_age_sigma} 
%\end{center}
%\end{figure}

Fig. \ref{Ca_Fe_versus_Na} shows the NaI2.21 index as a function of the $K$-band indices Ca {\scriptsize I} and Fe {\scriptsize I}, respectively, for the stars of the IRTF-library, our scaled-solar SSP models and the various galaxy samples. Compared to the older work by \citetalias{Silva08} and \citetalias{Marmol09}, we are able to display an increased sample of galaxies and a much larger set of comparison stars. We chose to visualize the NaI2.21 index here as a function of these two other $K$ band indices, since the galaxies of \citetalias{Silva08} and \citetalias{Marmol09} cover only this spectral range. Table \ref{indices_Table} lists the definitions of the various line strength indices presented in this paper.

\begin{figure}
\begin{center}
 \resizebox{\hsize}{!}{\includegraphics{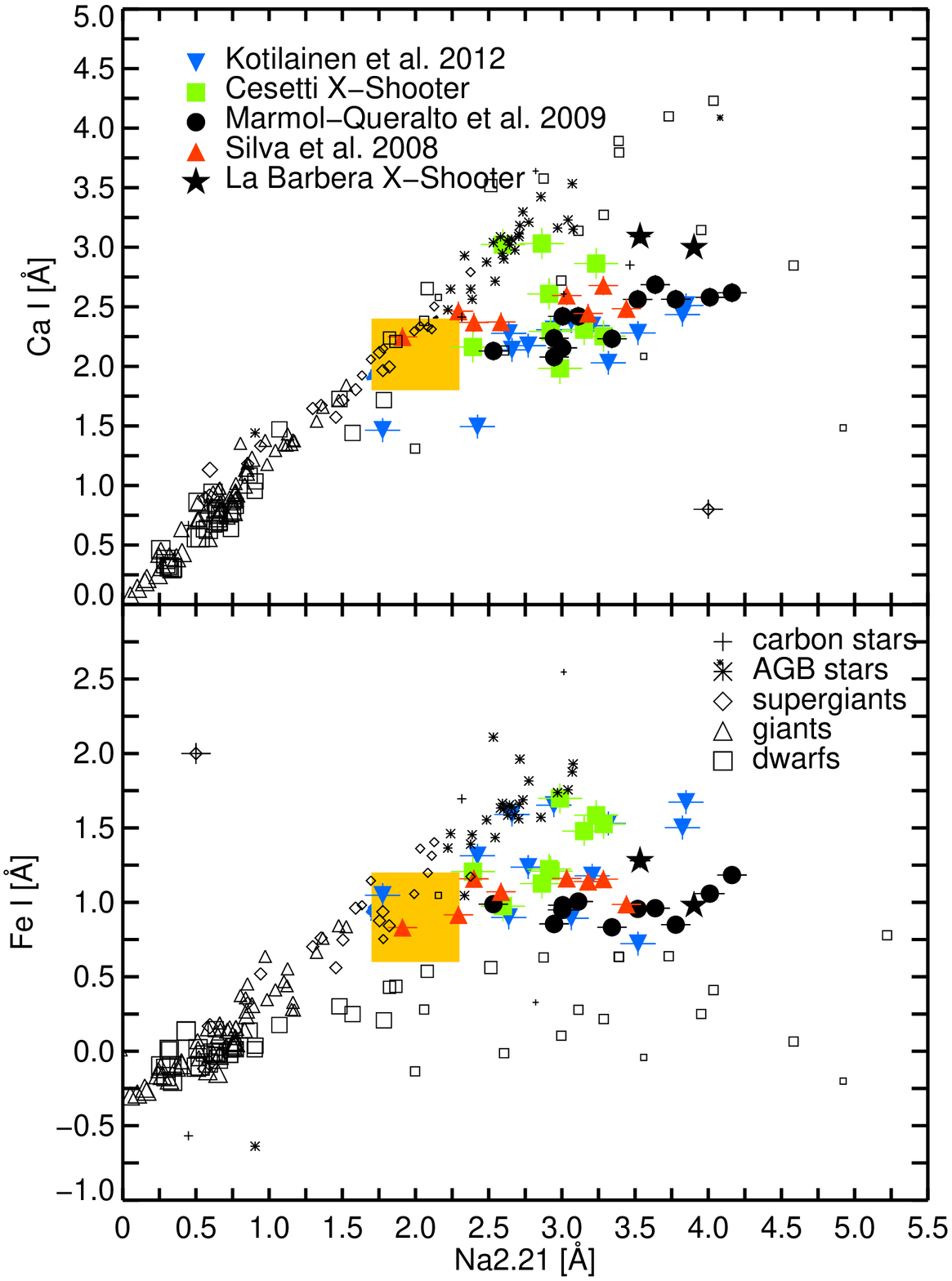}}
  \caption{\small{Upper panel: The Ca {\scriptsize I} index in the $K$-band as a function of the NaI2.21 index for the ETG samples of \citetalias{Silva08} (red triangles), \citetalias{Marmol09} (black dots), \citet[][, blue triangles upside down]{Kotilainen12}, Cesetti (green squares), the galaxies of \citet[][, black stars]{LaBarbera16a} as well as for 180 stars of the IRTF library. The different stellar types are illustrated with different symbols (pluses for carbon stars, asterisks for AGB stars, diamonds for supergiants, triangles for giants and squares for dwarfs) which grow in size according to their ${\rm \textit{T}_{eff}}$. The temperature range covered by the IRTF stars is ${\rm 2000 - 7500 \, K}$. The ocre-shaded squares show the position of the predictions of our scaled-solar SSP models of [Na/Fe] = 0. Errorbars indicate the typical errors in the measurements of the indices. Lower panel: As in the upper panel, but now for the Fe {\scriptsize I} index in the $K$-band as a function of the NaI2.21 index. All model, ETG and stellar spectra have been convolved to a common resolution of ${\rm 360 \, kms^{-1}}$ before measuring the respective indices.} }\qquad
 \label{Ca_Fe_versus_Na}
\end{center}
\end{figure}

In both the upper and the lower panel of Fig. \ref{Ca_Fe_versus_Na}, almost all IRTF stars trace a clearly defined, rather narrow sequence. Along these sequences, the line strength indices increase while the ${\rm \textit{T}_{eff}}$ of the stars decreases from hot giants and dwarfs of around ${\rm 6000 - 7000 \, K}$ to cool dwarfs and AGB stars of temperatures of $\sim 3000 -$ ${\rm 4000 \, K}$. The only stars which do not fall onto these sequences are a few of the lower temperature dwarfs in the case of Ca {\scriptsize I} as a function of NaI2.21. In the case of the Fe {\scriptsize I}, a large number of dwarfs do not follow this sequence. Dwarf stars deviate from these sequences of stars visible in Fig. \ref{Ca_Fe_versus_Na}, since they feature smaller Ca { \scriptsize I} or Fe { \scriptsize I} indices than on-sequence stars for a given value of the NaI2.21 index. Whereas the model predictions lie on the sequences delineated by the single IRTF stars, as one may expect, the galaxies follow a separate sequence which branches off at NaI2.21 values of between ${\rm \approx 2.0 \, \r{A}}$ and ${\rm \approx 2.5 \, \r{A}}$. Hence, only the off-sequence dwarf stars are able to trace fairly well the NaI2.21 and Ca {\scriptsize I} indices of most of the galaxies. However, these dwarf stars exhibit too small Fe {\scriptsize I} indices compared to the indices measured for the ETGs.

Considering the small contribution to the total light of the model spectra of those few, low-luminous dwarf stars, it is not surprising that the model predictions coincide with the central part of the sequences traced by the indices of the IRTF stars in Fig. \ref{Ca_Fe_versus_Na}. Hence, Fig. \ref{Ca_Fe_versus_Na} also shows clearly that it is impossible to reproduce the observed indices of the ETGs by changing the weighting of the IRTF stars in the models in any plausible way. 

We further calculated models based on a modified set of isochrones which are ${\rm 200 \, K}$ cooler at the tip of the red giant branch (RGB), as it might be possible that a too warm giant branch might be able to explain the too small NaI2.21 values obtained from the models compared to the observed ETGs. However, the models based on isochrones with a cooler RGB tip yield NaI2.21 values which are only around ${\rm 0.05 \, \AA}$ larger than the NaI2.21 values measured from the models with underlying standard isochrones. Hence, we conclude that the temperature of the upper RGB and AGB in the BaSTI and Padova isochrones is not an issue for our analysis of the NaI2.21 index.
%The lower panel of Fig. \ref{Ca_Fe_versus_Na} shows that the predictions of the models with a bottom-heavy IMF lead to smaller Fe {\scriptsize I} and slightly larger NaI2.21 values. This agrees well with what is expected from an enhanced contribution of dwarfs in Fig. \ref{Ca_Fe_versus_Na}.

Furthermore, Fig.\ref{Ca_Fe_versus_Na} gives a first hint that models of an enhanced AGB star contribution will not be able to reproduce the observed galaxies. Notice that the AGB stars follow the sequence of indices in both panels of Fig. \ref{Ca_Fe_versus_Na}, whereas the observed galaxies lie offset from them.  

%\begin{figure}[!t!]
%\begin{center}
% \resizebox{\hsize}{!}{\includegraphics{MgI_versus_NaI_letter.eps}}
%  \caption{As Fig. \ref{Ca_versus_Na}, now for the Mg {\scriptsize I} index at ${\rm 1.71 \, \mu m}$ as a function of the Na2.21 index.}\qquad
%   \label{Mg_versus_Na}
%\end{center}
%\end{figure}

\subsection{General behaviour of the NaI2.21 index}

\begin{figure*}
\begin{center}
\resizebox{\hsize}{!}{\includegraphics{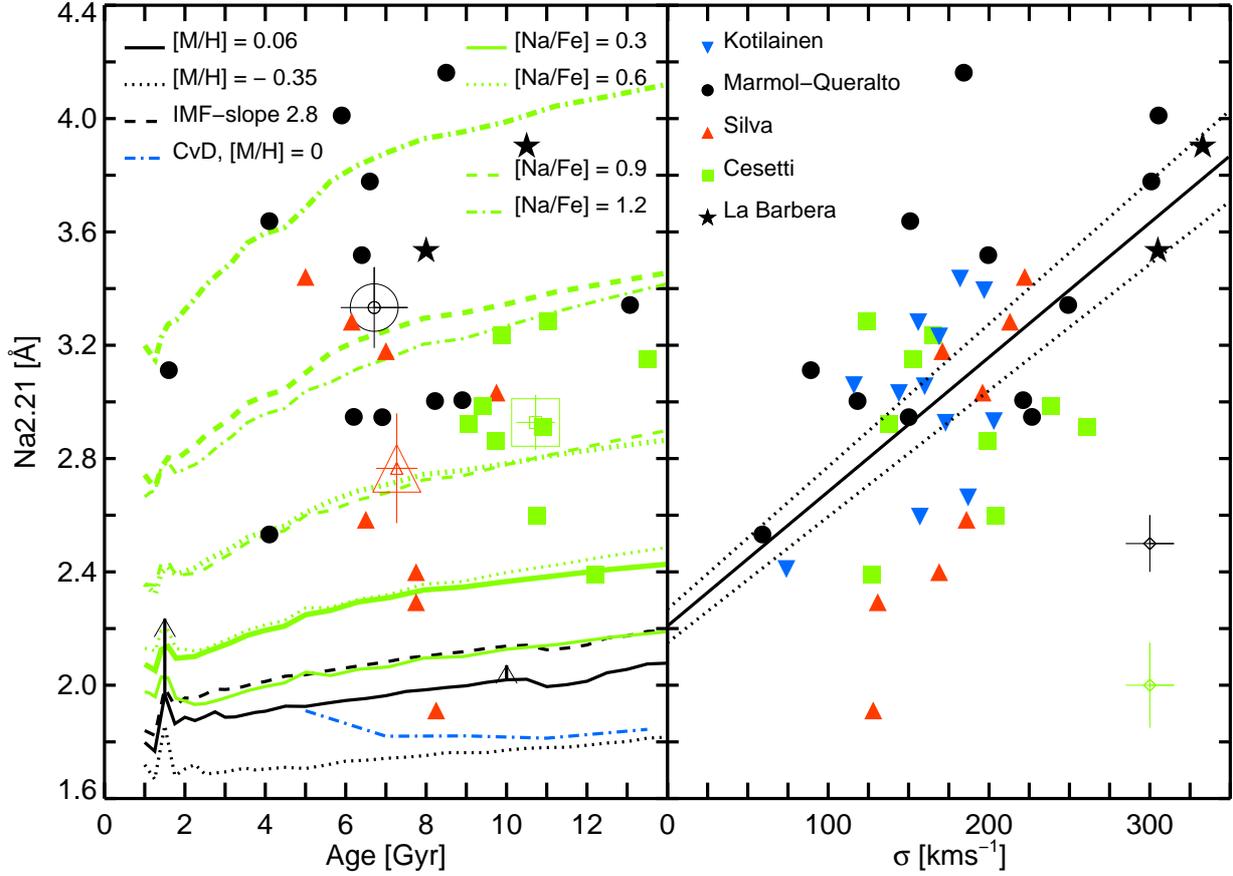}}
  \caption{\small{Left-hand panel: Behaviour of the NaI2.21 index as a function of age. The solid and the dotted black lines display the predictions from our (BaSTI-based) models of an underlying Kroupa-like IMF and of the metallicities ${\rm [M/H] = 0.06}$ and ${\rm [M/H] = - 0.35}$, respectively, whereas the dashed black line illustrates the respective index values measured from a SSP of Solar metallicity and bottom-heavy IMF of slope $\Gamma_{{\rm b}} = 2.8$ (see also legend). The dot-dashed blue line describes the predictions of the CvD models for Solar metallicity and a Kroupa-IMF. The green lines illustrate the NaI2.21 indices measured from various sets of our newly developed Na-enhanced models (see legend). The thin lines represent models of an underlying Kroupa-like IMF, whereas the thick lines describe the predictions of models based on a bottom-heavy IMF of slope $\Gamma_{{\rm b}} = 2.8$. Overplotted are the measurements of the same index for the early-type field galaxies of \citetalias{Marmol09} (black dots), of the Fornax cluster galaxies of \citetalias{Silva08} (red triangles), of the X-Shooter galaxies of Cesetti (green squares) and of \citet{LaBarbera16a} (black stars). For the first three samples, we also display their respective mean values including errors (large black circle, red triangle and green square, respectively). The black arrow at 1.5 Gyr marks the change in the NaI2.21 index when moving from our standard models to one with a significantly enhanced AGB star contribution of 70 \% of the total luminosity. The black arrow at 10 Gyr, indicates the change in the NaI2.21 index when adding a mass fraction of 3 \% of this young, AGB-enhanced component to the underlying standard SSP model of 10 Gyr. Right-hand panel: The NaI2.21 index as a function of the central velocity dispersion $\sigma$ for the same four samples of ETGs. The black solid line indicates the best weighted linear fit to all the displayed ETGs, whereas the two black dotted lines show the $1 \, \sigma$ confidence level. The two crosses delineate the typical errors in NaI2.21 and in $\sigma$ for the galaxies of \citetalias{Silva08}, \citetalias{Marmol09} and \citet{LaBarbera16a} (black cross) and Cesetti et al. (green cross), respectively. }}\qquad
\label{Na_age_sigma}
\end{center}
\end{figure*}

The left-hand panel of Fig. \ref{Na_age_sigma} displays the index values of the NaI2.21 determined from our SSP models of varying metallicities, IMFs and [Na/Fe] abundances and measured from our samples of galaxies as a function of age. Table \ref{indices_Table} contains the definition of the NaI2.21 index according to \citet{Frogel01}, which we use throughout this paper.

\begin{table*}
\caption{ Definitions of the line strength indices in the optical and in the $K$-band appearing in this paper.}
\label{indices_Table}
\centering
\begin{tabular}{l c c c c}

\hline
Index  & Blue pseudo-  &  Central  &  Red pseudo-  & Source\\
          & continuum [$\r{A}$] & bandpass [$\r{A}$] & continuum [$\r{A}$] &   \\
% & with the MILES library & with the IRTF library\\
\hline
C4668 & 4611.500 - 4630.250 & 4634.000 - 4720.250 & 4742.750 4756.500 &  \citet{Trager98} \\
NaD  & 5860.625 - 5875.625 & 5876.875 - 5909.375 &  5922.125 5948.125 & \citet{Trager98} \\
Na {\scriptsize I}(${\rm 2.21 \, \mu m}$) & 21910 - 21966 &  22040 - 22107 &  22125 - 22170 & \citet{Frogel01}\\
Ca {\scriptsize  I} (${\rm 2.26 \, \mu m}$)& 22450 - 22560 &  22577 - 22692 &  22700 - 22720 & \citet{Frogel01}\\
Fe {\scriptsize I} (${\rm 2.23 \, \mu m}$)& 22133 - 22176 &  22251 - 22332 &  22465 - 22560 &  \citet{Silva08}\\
Fe  {\scriptsize I} (${\rm 2.24 \, \mu m}$)& 22133 - 22176 &  22369 - 22435 &  22465 - 22560 &  \citet{Silva08}\\

  \hline
\end{tabular}
%\tablefoot{
%\medskip
\end{table*}

The ages of the galaxies of \citetalias{Marmol09} and of Cesetti were determined by \citet{Sanchez06b} using the ${\rm H \beta}$ index in the optical range and a preliminary version of the population synthesis models of \citet{Vazdekis10}. The ages of the galaxies of \citetalias{Silva08}, however, were computed by \citet{Kuntschner98} based on the models of \citet{Worthey94}. Fig. \ref{Na_age_sigma} shows that neither our base SSP models nor the CvD models are able to fit the NaI2.21 indices of most of the observed galaxies. Their index values lie in the range between 1.9 and more than ${\rm 4.3 \, \r{A}}$ and are thus enhanced compared to the values measured from both our scaled-solar and the CvD models. As Figure 16 in \citet{Meneses15b} shows, also choosing different isochrones does not have a large effect on the predictions for the NaI2.21. However,  Fig. \ref{Na_age_sigma} also shows that models of enhanced [Na/Fe] abundances yield significantly larger NaI2.21 indices compared to the scaled-solar ones, and that the former are in much better agreement with the observations. This is particularly true for models of steeper IMFs, on which the differential effect of the [Na/Fe] is larger. As an example, in Fig. \ref{Na_age_sigma}, we depict the set of models calculated for an underlying bottom-heavy IMF of slope $\Gamma_{{\rm b}} = 2.8$. For these models, a [Na/Fe]-abundance of 0.6 is sufficient to fit the mean NaI2.21 values of the galaxy samples of \citetalias{Silva08} and of Cesetti, while models of [Na/Fe]=0.9 are needed to reproduce the mean value of the galaxies of \citetalias{Marmol09}. Hence, we show for the first time that models with enhanced [Na/Fe] are actually able to match the large observed line strengths of NaI2.21.

Moreover, we see that galaxies located in the Fornax cluster tend to have, on average, smaller NaI2.21 line strengths than their counterparts in the field. We will further comment on this point in the section \ref{environment}.

Apart from a small peak for young ages between 1 and ${\rm 2 \, Gyr}$, the index values for the NaI2.21 predicted by our SSP models are almost constant as a function of age. However, the line strength indices increase by about ${\rm 0.3 \, \r{A}}$ for models of slightly supersolar metallicities (${\rm [M/H] \gtrsim 0.06}$) compared to subsolar ones of ${\rm [M/H]=-0.35}$. 

This behaviour of the NaI2.21 as a function of metallicity is consistent with the clear correlations between NaI2.21 and the optical metallicity indicators [MgFe] and C4668 found for the first time by \citetalias{Silva08} and \citetalias{Marmol09}, respectively. Both \citetalias{Silva08} and \citetalias{Marmol09} also noted the correlation between NaI2.21 and the central velocity dispersion $\sigma$ which we illustrate in the right-hand panel of Fig. \ref{Na_age_sigma}. We also show an error-weighted linear fit to all the galaxies including the $2 \sigma$ error and compute a Pearson correlation coefficient of 0.56. Such a dependence is expected because of the Mg$_2$--$\sigma$ relation \citep[e.g.][]{Colless99, Trager00, Kuntschner01} and since $\sigma$ correlates with metallicity \citep{Kuntschner00}. At the same time, the velocity dispersion $\sigma$ is also a proxy for the galaxy mass. However, NaI2.21 clearly correlates less well with $\sigma$ than Mg$_2$, and there is an offset visible between cluster and field galaxies \citepalias[see][]{Marmol09}. %Hence, we can deduce that the NaI2.21 absorption feature increases strongly for more massive and more metal-rich galaxies. 

Moreover, we constructed models of an enhanced contribution of AGB stars in order to study whether the predictions of such models result in a better fit to the observed values of the NaI2.21 index. For these models, we combined a base model spectrum of 1.5 Gyr, solar metallicity and Kroupa-like IMF, which was calculated excluding all AGB stars and the corresponding part of the isochrone with a model spectrum of the same parameters computed using only AGB stars. Our standard SSP model of 1.5 Gyr contains a bolometric luminosity fraction of 32 \% AGB stars. According to \citet{Maraston98}, at these ages, AGB stars can be however responsible for up to 70 \% of the total flux in the NIR. Hence, we changed the total weight in luminosity of the model made up of only AGB stars to 70 \% while reducing the weight of the model without AGB stars to 30 \%. As mentioned in \citet{Roeck15}, our IRTF-based stellar population models contain in total 32 AGB stars, which should be well representative for this population.

In the left-hand panel of Fig. \ref{Na_age_sigma}, the black arrow at 1.5 Gyr indicates the effect of such an AGB-enhanced model on the NaI2.21 index. However, since most of the galaxies have older mean ages, we also assumed a more realistic scenario in which this young, AGB-enhanced model constitutes just 3 \% in total mass on top of an old component of, e.g., 10 Gyr. According to \citet{Yi05} and \citet{Schiavon07}, only between 1 - 2 \% and at maximum 10 \% of the total mass of ETGs exists in the form of a young stellar component. The black arrow at 10 Gyr shows the very small change in the NaI2.21 when adding such a young, AGB-enhanced component of just 3 \% in total mass to the underlying old one. Results of a recent analysis by \citet{Vazdekis16} of colours and line strengths in the ultra-violet based on newly developed stellar population models suggest that this young component might be even smaller and make up for only around 0.1 - ${\rm 0.5 \, \%}$ in mass and might also be significantly younger ${\rm 0.1- 0.5 Gyr}$, consistent with residual star formation.

Interestingly, we find that both black arrows in Fig. \ref{Na_age_sigma} have only a very limited effect on the model predictions and do not improve significantly the fit of the model NaI2.21 indices to those of the observed galaxies. Hence, we are able to show clearly for the first time that the mismatch between models and observed galaxies cannot be explained by increasing the contribution of AGB stars to our models.

%In the middle panel of Fig. \ref{NaI_age_C_sigma}, we show the remarkable correlation between NaI2.21 and C4668 for all selected data samples.

Furthermore, models of a more bottom-heavy bimodal IMF of slope ${\rm \Gamma = 2.8}$ yield indices which are enhanced by ${\rm \approx 0.2 \, \r{A}}$ with respect to models of a standard Kroupa-like IMF. Such a bimodal IMF is characterized by a constant slope at very low masses of $ m < 0.6 M_{\odot}$, while the slope for higher masses is described by ${\rm \Gamma = 2.8}$ \citep[for more details about its exact form see][]{Vazdekis03, Vazdekis15}. \citet{LaBarbera13, Spiniello14} found a correlation between the slope of the IMF and the central velocity dispersion $\sigma$ of galaxies, according to which a slope of${\rm \Gamma = 2.8}$ corresponds to galaxies of typical around ${\rm \sigma \sim 300 kms^{-1}}$.

\citet{Spiniello12} and \citet{Jeong13} already reported the same effect for the NaD in the optical. However, these authors also showed that the rise in the index values observed for a bottom-heavy IMF alone is not sufficient to fully explain the large values of the NaD, in particular for massive systems with ${\rm \sigma > 250 \, km \, s^{-1}}$. On the other hand, the NaI1.14 index seems to be significantly more sensitive to the form of the chosen IMF \citep{Smith15}.

Our work shows, for the first time, that the same conclusion applies to the NaI2.21 feature, whose observed index values in early-type galaxies are too large to be reproduced correctly even by models with a bottom-heavy IMF. However, as mentioned before, if we combine a bottom-heavy IMF with significant [Na/Fe]-overabundances of 0.6 -- 0.9 dex, we are in fact able to fit most of the observed galaxies. It should be pointed out that a steep IMF together with a sodium enhancement of ${\rm [Na/Fe] \approx 0.6}$ is also very similar to the best fit found for the NaI1.14 index by \citet{Smith15}. However, their study is based on a different form of the underlying IMF \citep[see][]{Smith15}.

\subsection{Possible drivers of the NaI2.21 index}\label{indices_Na_drivers}

As mentioned in Section \ref{introduction}, the NIR is barely hampered by ISM and dust absorption. Thus, a scenario like the one mentioned by \citet{Jeong13} for the NaD in the optical which tries to explain the large index values by contamination due to interstellar absorption can be ruled out for the NaI2.21 in the $K$-band \citep[see also][]{Spiniello12}.

\begin{figure*}
\begin{center}
 \resizebox{\hsize}{!}{\includegraphics{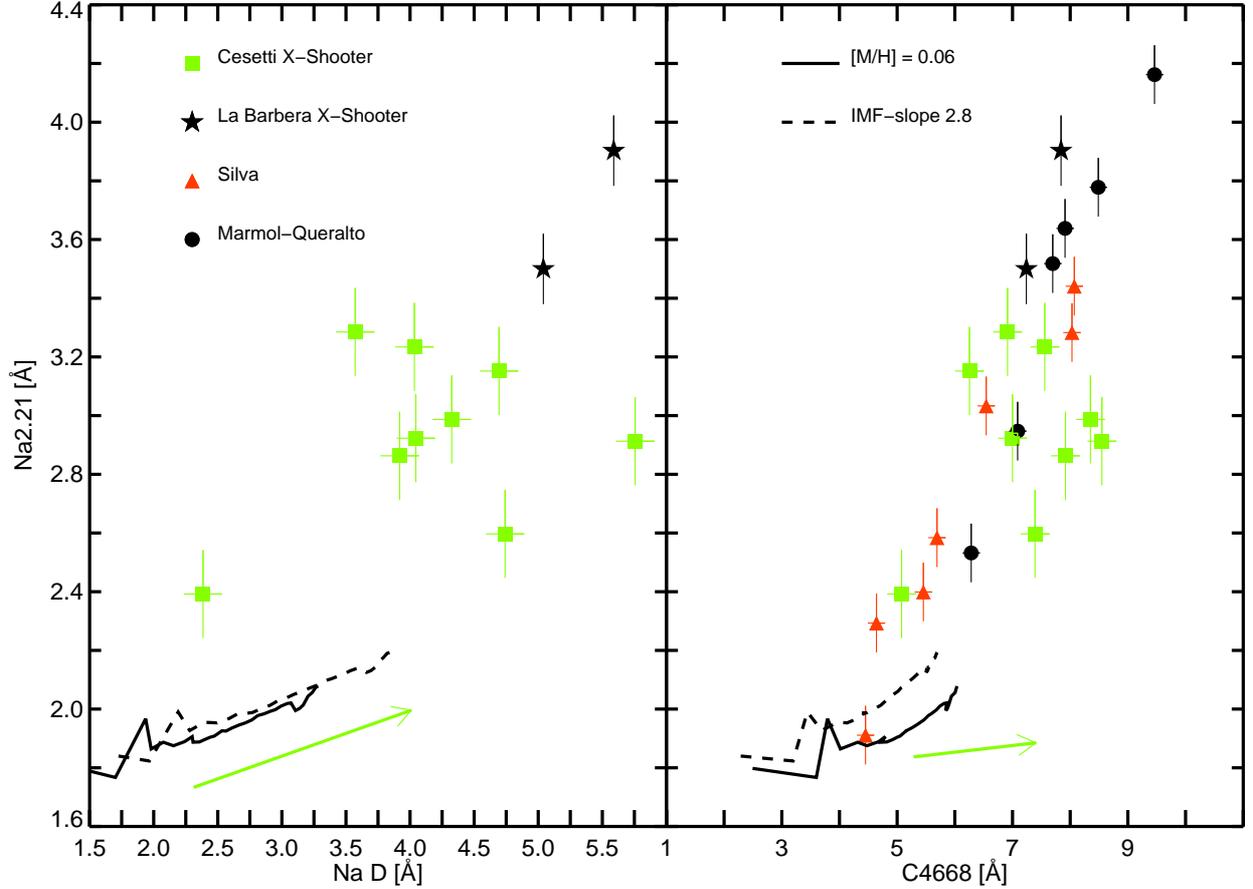}}
  \caption{\small{Left-hand panel: The values of the NaI2.21 index as a function of those of the optical NaD index at ${\rm 0.59 \, \mu m}$ for the galaxies of Cesetti (green squares) and those of \citet[][, black stars]{LaBarbera16a}. Moreover, the predictions of our SSP-models of Solar metallicities and a Kroupa-like IMF and of a bottom-heavy IMF, respectively, are overplotted (see also legend). The green arrow marks the change in the predictions of the CvD models when moving from ${\rm [Na/Fe]=-0.3}$ to ${\rm [Na/Fe]=0.3}$. Right-hand panel: The NaI2.21 indices of the galaxies of Cesetti as well as of subsets of the samples of \citetalias{Silva08} and \citetalias{Marmol09} as a function of the C4668 index in the optical. Again, also various model predictions are shown together with the predictions of the CvD models when moving from ${\rm [C/Fe]=-0.15}$ to ${\rm [C/Fe]=0.15}$ (green arrow). }}\qquad
 \label{Na_versus_C_NaD}
\end{center}
\end{figure*}

This leaves us with the explanation that most of our galaxies are Na-enhanced, while most of our stars are not. As we have seen in the left-hand panel of Fig. \ref{Na_age_sigma}, our sets of models which take Na-enhancement into account indeed yield larger NaI2.21-indices.
For the X-Shooter galaxies by Cesetti et al. and \citet{LaBarbera16a}, we can also measure the NaD index in the optical range which has been found to be sensitive to a significant extent to Na-enhancement \citep{Conroy12}. The left-hand panel of Fig. \ref{Na_versus_C_NaD} shows the behaviour of the NaI2.21 index as a function of the NaD index in the optical for those galaxies together with the predictions from our models. Quite surprisingly, we do not observe a clear correlation between the two indices (Pearson-coefficient of only ${\rm P=0.55}$). Also the measured values for the NaD index are quite high. According to \citet{Bica86}, NaD is also absorbed by the Milky Way. However, since all of our galaxies are located away from the Galactic plane, we consider the effect of Milky Way absorption to be rather small on our galaxies. According to Table 2 in \citet{Bica86}, the effect of the Milky Way absorption on the NaD of galaxies is only ${\rm \approx 0.1 \, \AA}$, if their reddening is smaller than ${\rm E(B-V) < 0.2}$. Based on the updated reddening maps of \citet{Schlafly11}, we estimated the reddening on all of our studied ETGs and found a maximum value of only ${\rm E(B-V) \approx 0.08}$. Hence, we can safely exclude a significant effect of Milky Way absorption on the NaD of our ETGs. As the right-hand panel of Fig. \ref{Na_versus_C_NaD} shows, a clear correlation is also absent between the NaI2.21 index and the optical C4668 index for the galaxies of Cesetti (${\rm P=0.35}$). However, in the case of the galaxies of \citetalias{Silva08} and \citetalias{Marmol09}, the NaI2.21 index clearly correlates with the optical C4668 index (${\rm P=0.95}$). This behaviour has been reported before by \citetalias{Marmol09} based on the same galaxies which we also plotted in Fig. \ref{Na_versus_C_NaD}.  We ascribe the absence of a clear correlation in the case of the galaxies of Cesetti to the fact that their indices are plagued by larger errors compared to the galaxies of the samples of \citetalias{Silva08} and \citetalias{Marmol09} which have probably been introduced by an imperfect telluric correction.%However, this correlation is less clear when considering only the Cesetti galaxies. 

.

We further assessed the question what element abundances are responsible for the absorption feature seen at ${\rm 2.21 \, \mu m}$ based on the theoretical stellar spectra and the CvD models, both of which are available for varying abundance ratios. Fig. \ref{Na_versus_Ca_Carlos} is equal to the upper panel of Fig. \ref{Ca_Fe_versus_Na}, apart from the IRTF stars which are not displayed for better visibility. Instead, we show the indices measured for the solar and the carbon-enhanced theoretical stellar spectra and those measured from various Na-enhanced models. With the exception of the two low-temperature dwarf stars with parameters $T_{{\rm eff}}=3500 \, K$, log(\textit{g})=5.0 and $ T_{{\rm eff}}=4000 \, K$, log(\textit{g})=4.7, all the other giant, supergiant and AGB stars fall onto the sequence traced by most of the IRTF stars in the upper panel of Fig. \ref{Ca_Fe_versus_Na}. Hence, only those two cool dwarf stars are able to reproduce the large NaI2.21 values observed for the ETGs. Already for Solar metallicities and no enhancement in any element, the two cool dwarf stars exhibit much stronger NaI2.21 indices than all other types of stars. The thin blue arrows in Fig. \ref{Na_versus_Ca_Carlos} mark the increase caused by an enhanced [Na/Fe] ratio. The cooler of the two dwarfs shows both a larger index of NaI2.21 at Solar metallicity and a by almost a factor of two stronger rise with Na enhancement compared to its ${\rm 500 \, K}$ hotter counterpart. Also carbon enhancement leads to a larger NaI2.21 value in the case of the cooler dwarf star (see the thin green arrow in Fig. \ref{Na_versus_Ca_Carlos}).For the slightly hotter one, however, the NaI2.21 does not change at all when increasing the [C/Fe] ratio from 0 to ${\rm 0.2 \, dex}$. 

\begin{figure*}
\begin{center}
 \resizebox{\hsize}{!}{\includegraphics{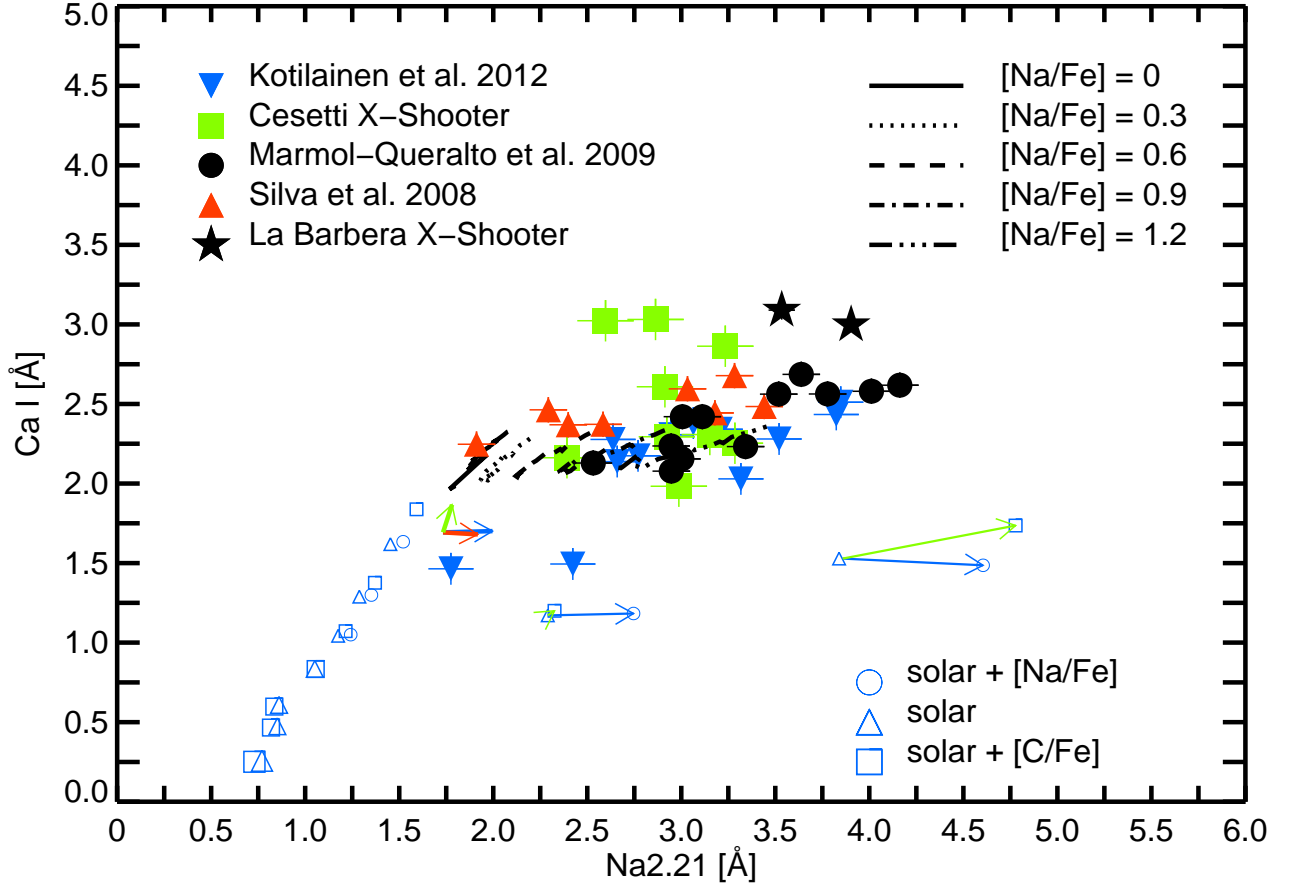}}
\caption{\small{As the upper panel of Fig. \ref{Ca_Fe_versus_Na}, but now without the IRTF-stars. Instead, we show the predictions of our scaled-solar models of ${\rm [M/H] = 0.06}$ (solid line) and of our [Na/Fe]-enhanced models based on a Kroupa-like IMF of slope $\Gamma_{{\rm b}} = 1.3$ (see legend). For all the models, the parameter age is varying between 1.5 and 17 Gyr along the black lines. Moreover, we display the sets of Solar (blue triangles), Na-enhanced (${\rm [Na/Fe]=0.2 \, dex}$, blue circles) and Carbon-enhanced (${\rm [C/Fe]=0.2 \, dex}$, blue squares) theoretical stellar spectra. Like in Fig. \ref{Ca_Fe_versus_Na}, the symbols denoting these theoretical stellar spectra grow in size according to their ${\rm \textit{T}_{eff}}$ in the range ${\rm 2000 - 7500 \, K}$. The thin green arrows between the stars denote the effect of Carbon enhancement, while the thin blue errors between the stars denote the effect of Na-enhancement. The thick blue, green and red arrows which share the same starting point exhibit the effects of Na, Carbon and Si enhancement (+0.3, +0.15 and +0.3, respectively) on the two displayed indices according to the CvD-models with a Kroupa IMF. }}\qquad
 \label{Na_versus_Ca_Carlos}
\end{center}
\end{figure*}

In Fig. \ref{Na_versus_Ca_Carlos}, the thick blue arrow indicating the effect of increasing Na abundances in the CvD models on the indices of NaI2.21 points in the correct direction towards the values measured for the ETGs. However, for an enhancement of ${\rm \Delta [Na/Fe] = 0.6 \, \AA}$, the CvD models predict a rise of the NaI2.21 index of only ${\rm \Delta (NaI2.21) = 0.26 \, \AA}$. Such an increase is too small to fit the strong NaI2.21 features observed in many galaxies. On the contrary here, as the solid and dashed black lines in Fig. \ref{Na_versus_Ca_Carlos} show, in the case of our new Na-enhanced models with a Kroupa-like IMF, a same enhancement of ${\rm \Delta [Na/Fe] = 0.6 \, \AA}$ leads to almost double the increase of the NaI2.21 index compared to the CvD models.

%Since we have seen from the behaviour of the theoretical stellar spectra that the effect of [Na/Fe] enhancement is much more pronounced for dwarf stars than for all other types of stars, Na-enhanced models of an underlying bottom-heavy IMF most likely would give a larger increase in the EWs of the NaI2.21.

Moreover, also CvD models with supersolar [Si/Fe] abundance ratios cause an increase in the strength of the NaI2.21 feature (see thick red arrow in Fig. \ref{Na_versus_Ca_Carlos}). This is expected, since Si {\scriptsize I} is a significant absorber in the region of the NaI2.21 \citepalias{Silva08}. However, according to \citet{Conroy14}, the [Si/Fe] ratio does not exceed 0.2 dex in most massive galaxies. Therefore, the supersolar [Si/Fe] abundance should have only a limited effect on the NaI2.21 index.

The thick green arrow marks the almost negligible impact of carbon-enhanced CvD models on the indices. This behaviour could be explained by the fact that in our analysis of the theoretical stellar spectra, [C/Fe] enhancement affects only the coolest dwarf star. Thus, its combined effect on the models is most likely significantly smaller than that of [Na/Fe] enhancement. %Furthermore, we cannot exclude that the theoretical modelling underestimates the effect of an enhanced [C/Fe] ratio on these coolest dwarf stars.

Since CvD models of varying abundance ratios are computed only for a Kroupa-like IMF, we cannot use these models to analyze whether the NaI2.21 index might be driven by enhanced [Na/Fe] and [C/Fe] ratios together with a bottom-heavy IMF according to the individual velocity dispersions of the galaxies \citep[see, e.g.,][]{LaBarbera13, Ferreras13, Spiniello14} and La Barbera et al., 2016b. With our new models, it is however possible to take both larger [Na/Fe] ratios up to ${\rm [Na/Fe]=1.2}$ and additionally a bottom-heavy IMF into account. The combined effect of those two factors indeed results in much larger  NaI2.21 indices, roughly in the range of what is observed for the ETGs. Moreover, from the theoretical stellar spectra, we saw that very cool, low mass dwarfs show a strong increase in the NaI2.21 index as a function of carbon enhancement and for two of our samples of observed galaxies \citepalias{Silva08, Marmol09}, the NaI2.21 index is clearly correlated with the [C/Fe] ratio. Therefore, we speculate that a likely scenario to explain the large values observed for the NaI2.21 index in many of the examined galaxies combines the effects of [Na/Fe] enhancement and of a bottom-heavy IMF with a smaller impact of also [C/Fe] and even to some extent [Si/Fe] enhancement. Notice that by considering a bottom-heavy IMF and the additional contribution of the [C/Fe] and [Si/Fe] enhancement to the strength of the NaI2.21 absorption feature, we do not need to consider such high Na-abundances of larger than ${\rm [Na/Fe] \sim 0.9}$ in order to fit most ETGs. However, we cannot exclude the possibility that there exists an unknown effect which is responsible for both the strong NaI2.21 values and for the pronounced C4668 features. 

Recently, \citet{Smith15b} claimed that the combined effect of a very bottom-heavy IMF and of [Na/Fe] enhancement is needed to explain the large NaI1.14 indices observed for massive ETGs based on the theoretical CvD models. However, they also mention that both a full spectrum fit and a number of high-mass galaxies are inconsistent with such a very bottom-heavy IMF.

\subsection{The role of galactic environment}\label{environment}

Comparing the mean values of NaI2.21 for the two samples of \citetalias{Marmol09} (field galaxies) and \citetalias{Silva08} (Fornax cluster galaxies), we measured a difference of $\sim 2.4 \sigma$ (see left-hand panel of Fig. \ref{Na_age_sigma}). Since both the \citetalias{Silva08} and \citetalias{Marmol09} data were acquired with the same instrument, and refer to the aperture size, this difference should not be affected by systematics. We indeed obtain the same index values (with differences ${\rm \Delta(NaI2.21)< 0.005 \, \AA}$ for the two galaxies NGC 1379 and NGC 1404, which have been re-observed by \citetalias{Marmol09} in order to be able to confirm reproducibility of the measurements of \citetalias{Silva08}. Hence, our result confirms the dichotomy between the two environments, which has been previously reported by \citetalias{Marmol09}. Interestingly, the Cesetti-galaxies, which are a mixture of field, group and cluster galaxies, lie in between the samples of \citetalias{Silva08} and \citetalias{Marmol09}.

We speculate that differences in the [C/Fe] abundances between field and cluster galaxies could also be at the origin of the observed dichotomy between the NaI2.21 values of the ETGs depending on their galactic environment. According to \citet{Sanchez03, Carretero04, Carretero07}, in denser environments like clusters, star formation happens on shorter timescales than in the field. Therefore, the relatively lower carbon in cluster galaxies compared to their counterparts in the field could be explained by the fact that the former ones had been formed before carbon was massively ejected into the interstellar medium by stars of intermediate masses. Furthermore, \citet{Carretero07} found that the duration of star formation depends less on the environment for galaxies of higher masses than for those of intermediate and low masses. In fact, the dichotomy which we observe for the NaI2.21 index seems to be particularly pronounced for galaxies with ${\rm \sigma < 200 kms^{-1}}$ (see right-hand panel of Fig. \ref{Na_age_sigma}), which have the smallest masses.

While it is feasible to fit the mean NaI2.21 value of the cluster galaxies of \citetalias{Silva08} with models of a bottom-heavy IMF and ${\rm [Na/Fe]=0.6}$, a larger [Na/Fe]-enhancement of around ${\rm [Na/Fe]=0.9}$ is needed to reproduce the mean NaI2.21 of the field galaxies of \citetalias{Marmol09}. Instead of assuming such a very high [Na/Fe]-ratio, a larger Carbon abundance in the field galaxies than in the cluster ones could also account for this difference.

%It should be however pointed out that these conclusions apply to the NaI2.21 index as it is defined in this work (see Table \ref{indices_Table}. We cannot exclude that different definitions of the central passband and of the pseudocontinua change the reported sensitivity of the NaI2.21 to the [Na/Fe] or [C/Fe] abundance ratios. Moreover, it would be also possible that the sensitivity of the dwarf stars to [C/Fe] is mainly related to the pseudocontinua instead of to the actual NaI2.21 feature. 

A similar explanation was made by \citet{Spiniello12} and \citet{Jeong13} for the NaD index in the optical range. The latter were able to reproduce the observed large indices of the NaD based on models which combine the effects of [Na/Fe], $\alpha$ and [M/H] enhancement with a more bottom-heavy IMF.

\section{Conclusion}\label{conclusion}

In accordance with earlier work \citep{Silva08, Marmol09, Cesetti09}, the values which we measure for the NaI2.21 index in the $K$-band for various samples of ETGs are all significantly larger than those we obtain from our scaled-solar models of varying ages, metallicities and IMFs. While absorption due to interstellar gas and dust does not play an important role in the $K$ band, we can understand this behaviour when measuring also the NaI2.21 indices of the single stars of the IRTF library, of other theoretical stars and of [Na/Fe]-enhanced models. In a diagram of the NaI2.21 index as a function of other indices in the $K$-band, most of the stars form a clear sequence which is offset from the location of most of the galaxies. Only cool dwarf stars trace the location of the ETGs in such an index-index diagram. In fact, we show that a [Na/Fe]-enhanced version of our SSP models based on an underlying more bottom-heavy IMF is able to improve the situation significantly by yielding larger index values for the NaI2.21.

Moreover, for two of our studied samples of galaxies, we see a correlation between the values of the NaI2.21 index and the [C/Fe] abundance ratios. We also show that, although AGB stars have often been claimed to be an important ingredient to explain the observed properties of ETGs in the NIR spectral range \citep[see, e.g.,][]{Maraston09, Meneses15b}, an enhanced contribution of AGB stars to the models does not provide any significantly better fit to the NaI2.21 indices observed in ETGs. 

Consequently, we propose a scenario in which the main discrepancy between the NaI2.21 indices measured from our standard, solar-scaled stellar population models and those of the observed ETGs can be explained by assuming [Na/Fe]-enhancement combined with a bottom-heavy IMF. We ascribe the remaining smaller difference to the effect of [C/Fe] and maybe [Si/Fe] enhancement. 

Inspired by the work of \citet{Sanchez03} and \citet{Carretero04} and the fact that the NaI2.21 index seems to be affected to some extent by the [C/Fe] abundance ratio, we speculate that the dichotomy between NaI2.21 values of field and cluster ETGs could be caused by differences in the carbon abundances as a function of the environment in which galaxies reside.

\section{Acknowledgements}

We thank the referees, Dr. R. Smith and Dr. G. Worthey, for their critical yet stimulating comments. Part of this work involves observations made with ESO Telescopes at the Paranal Observatory under programmes ID 092.B-0378 and 094.B-0747 (PI: FLB). We acknowledge financial support to the DAGAL network from the People Programme (Marie Curie Actions) of the European Union\textquotesingle s Seventh Framework Programme FP7/2007-2013/ under REA grant agreement number PITN-GA-2011-289313, from the European Union's Horizon 2020 research and innovation programme under Marie Sk\l{}odowska-Curie grant agreement No 721463 to the SUNDIAL ITN network, and from the Spanish Ministry of Economy and Competitiveness (MINECO) under grant number AYA2013-41243-P, AYA2013-Y8226-C3-1-P and AYA2016-76219-P.

\bibliography{bibliography}

\appendix

\section{Parameters and indices of the observed galaxies}

In Tables \ref{measurements_I_Table} and \ref{measurements_II_Table}, we present the measured line strength indices as well as other relevant parameters of the ETGs studied in this paper

\begin{table*}
\caption{ List of all ETGs studied in this paper. Column 1 gives their names, while column 2 contains their central velocity dispersions $\sigma$ and column 3 their ages. In columns 4 - 7, we indicate the $K$ band line strength indices NaI2.21, Ca {\scriptsize I}, Fe {\scriptsize I} a and Fe {\scriptsize I} b (which together make up the combined index Fe {\scriptsize I}), which we have measured at a resolution of ${\rm \sigma = 360 \, kms^{-1}}$. The authors which have observed the various galaxies are mentioned in column 8.}
\label{measurements_I_Table}
\centering
\begin{tabular}{l c c c c c c c }

\hline
name of  & $\sigma$     & age      & NaI2.21   & Ca {\scriptsize I}        & Fe {\scriptsize I} a   & Fe {\scriptsize I} b & observed \\
galaxy     & [$kms^{-1}$]  & [Gyr]  &  [$\AA$] & $[\AA]$                     & [$\AA$]                 &   [$\AA$]             &         by    \\
% & with the MILES library & with the IRTF library\\
\hline

NGC1344   &   171     &    7.0  &  3.179 &  2.444 &  1.369 &  0.911  & \citetalias{Silva08}\\
NGC1374    &  196     &    9.8  &  3.033 &  2.595 &  1.458 &  0.863   & \citetalias{Silva08}\\
NGC1379   &  130.8    &   7.8  &  2.293 &  2.463  & 1.027 &  0.805    & \citetalias{Silva08}\\
NGC1380   &    213      &   6.2 &  3.283 &  2.678  & 1.399  & 0.913   & \citetalias{Silva08}\\
NGC1381   &   169        &  7.8  & 2.399  & 2.369  & 1.346  & 0.971   & \citetalias{Silva08}\\
NGC1404  &     222.2    &  5.0  &  3.441 &  2.484 &  1.336 &  0.637  & \citetalias{Silva08}\\
NGC1419  &     128       &  8.3  &  1.911  & 2.247  & 1.023 &  0.639   & \citetalias{Silva08}\\
NGC1427   &    186       &   6.5 &  2.584 &  2.372 &  1.331 &  0.811   & \citetalias{Silva08}\\

NGC3605   &      59.1          &   4.1 &       2.532  & 2.129 &  1.297  & 0.679    & \citetalias{Marmol09}\\
NGC3818   &       199.5       &   6.4 &        3.518 &  2.562 &  1.327 &  0.585  & \citetalias{Marmol09}\\
NGC4261   &       300.9       &   6.6 &        3.778  & 2.563 &  1.180 &  0.520   & \citetalias{Marmol09}\\
NGC4564   &       184.3       &   8.5 &        4.162  & 2.618 &  1.556  & 0.811  & \citetalias{Marmol09}\\
NGC4636   &       221.2       &    8.9 &        3.006 &  2.420 &  1.275 &  0.687  & \citetalias{Marmol09}\\
NGC4742   &       89.2         &    1.6 &       3.112  &  2.420 &  1.286  & 0.725   & \citetalias{Marmol09}\\
NGC5796   &       305.4        &   5.9 &        4.011  & 2.580 &  1.339  & 0.778   & \citetalias{Marmol09}\\
NGC5813   &       226.7        &   6.2 &        2.947 &  2.078 &  1.274  & 0.437   & \citetalias{Marmol09}\\
NGC5831   &       151.0        &   4.1 &        3.638 &  2.686  & 1.458  & 0.464   & \citetalias{Marmol09}\\
ESO382-G016  &    249.2     &  13.1 &           3.342 &  2.232 &  1.193 &  0.472   & \citetalias{Marmol09}\\
ESO446-G049  &    118.2     &   8.2 &         3.003  & 2.155  & 1.076  & 0.823   & \citetalias{Marmol09}\\
ESO503-G012  &    150.1     &   6.9 &          2.946 &  2.236  & 1.195 &  0.514  & \citetalias{Marmol09}\\

IC4214      & 173     & - &  3.317  & 2.029  & 1.844 & 1.215    & \citet{Kotilainen12}\\
NGC2380  &    187 &  - &  3.520  & 2.280 &  0.638 & 0.808    & \citet{Kotilainen12}\\
NGC2613  &    169  & - &  2.423  & 1.494 &  1.799 & 0.827    & \citet{Kotilainen12}\\
NGC2781  &    144 &  - &  2.637  & 2.277 &  1.282 & 0.516   & \citet{Kotilainen12}\\
NGC3056  &     74  &  - & 1.775   & 1.464 &  1.286 & 0.810   & \citet{Kotilainen12}\\
NGC3892  &    116  & - &  2.945  & 2.307 &  1.756  &1.549   & \citet{Kotilainen12}\\
NGC4179  &    157 &  - &  3.824  & 2.434 &  1.447 & 1.557   & \citet{Kotilainen12}\\
NGC4546  &    197  & - &  3.064  & 2.380 &  1.076 & 0.710   & \citet{Kotilainen12}\\
NGC4856  &    160  & - &  2.659  & 2.138 &  1.859 & 1.322   & \citet{Kotilainen12}\\
NGC4958  &    156  & - &  2.770  & 2.174 &  1.496 & 0.977   & \citet{Kotilainen12}\\
NGC5101  &    203 & -  &  3.848  & 2.511  & 1.899  & 1.447   & \citet{Kotilainen12}\\
NGC5507  &    182 & -  &  3.207  & 2.341 &  1.025 & 1.332  & \citet{Kotilainen12}\\

NGC3377     &          137.5  & 9.05  &  2.923  &   2.294  &   1.419  & 1.001  & Cesetti et al. \\
NGC3379     &          204.2  & 10.8 &  2.597  &   3.023  &   1.018  & 0.930  & Cesetti et al. \\
NGC584       &         199.3   & 9.7 &  2.863  &   3.031  &   1.083  & 1.169   & Cesetti et al. \\
NGC636       &         165.1   & 9.9 &  3.234  &   2.863  &   1.812  & 1.359   & Cesetti et al. \\
NGC1425   &           126.9 &  12.2&  2.392  &   2.162  &   1.510  & 0.903   & Cesetti et al. \\
NGC3115      &       261.1    & 10.9 &  2.913  &   2.608  &   1.381  & 1.065    & Cesetti et al. \\
NGC1700     &        238.7    & 9.4 &  2.987  &   1.983  &   1.832  & 1.564    & Cesetti et al. \\
NGC1357     &        124.0    &  11.0&  3.285  &   2.253  &   1.723  & 1.329    & Cesetti et al. \\
NGC2613    &         152.9   & 13.5 &  3.152 & 2.308 &  2.052  & 0.906 & Cesetti et al. \\

XSG1     &     332.8 & 10.5 &    3.903 & 2.998 & 1.118 & 0.842 & \citet{LaBarbera16a} \\
XSG2      &      305   &  8 &   3.534  & 3.09  & 0.729  & 1.830 & \citet{LaBarbera16a} \\

  \hline
\end{tabular}
%\tablefoot{
%\medskip
\end{table*} 

\begin{table*}
\caption{ Abundance ratios and optical indices for some of the ETGs studied in the paper (when available). Column 1 gives the names of the galaxies, while column 2 contains their total metallicities. In columns 3 - 5, we indicate their [Mg/Fe], [C/Fe] and [Na/Fe] abundance ratios, respectively. Column 6 and 7 contain the measurements for the two optical indices C4668 and NaD. The authors which have observed the various galaxies are mentioned in column 8.}
\label{measurements_II_Table}
\centering
\begin{tabular}{l c c c c c c c c}

\hline
name of      &  [M/H]  &  [Mg/Fe]  & [C/Fe]  &  [Na/Fe]& C4668  & NaD        & observed\\
galaxy        &             &                 &            &              &  [$\AA$] &               &   by       \\
% & with the MILES library & with the IRTF library\\
\hline

NGC1374  &    0.1   &  0.23   &  0.1304      &  - &  6.544  & - &\citetalias{Silva08} \\
NGC1379  &    0.09 &    0.23 &  -0.02   &  - &      4.641  & - &\citetalias{Silva08} \\
NGC1380  &   0.285    & 0.18    &  0.117      &  - &   8.031  & - & \citetalias{Silva08} \\
NGC1381  &   0.125    & 0.18  & -0.0399      &  - &    5.455  & - & \citetalias{Silva08} \\
NGC1404  &    0.31   &  0.19   &  0.1042       & - &    8.071  & - & \citetalias{Silva08} \\
NGC1419  &     -0.0375  &   0.19 &  -0.0255  & - &         4.451  & - &  \citetalias{Silva08} \\
NGC1427  &    0.125  &   0.18 &  -0.0124     &  - &     5.691  & - & \citetalias{Silva08} \\
NGC3605  &    0.15  &  0.09 &  -0.0134    &  - &     6.284   & - & \citetalias{Marmol09} \\
NGC3818  &    0.22  &   0.23  &   0.1521      &  - &    7.697   & - & \citetalias{Marmol09} \\
NGC4261  &    0.29  &   0.28  &   0.216     &  - &  8.487    & - & \citetalias{Marmol09} \\
NGC4564  &    0.24  &   0.2  &   0.3318     &  - &    9.462    & - & \citetalias{Marmol09} \\
NGC5813  &   0.27   &  0.3 &   0.0644     &  - &     7.089    &- &  \citetalias{Marmol09} \\
NGC5831  &   0.3   &  0.21 &   0.0974     &  - &     7.911    & - & \citetalias{Marmol09} \\

NGC3377  &   0.1312  &  0.2261 &   0.1626  &    0.3705 & 7.000    &   4.045 & Cesetti et al. \\
NGC3379  &   0.2312  &  0.2798 &   0.1225  &  0.4283  & 7.390     &     4.741 & Cesetti et al. \\
NGC584    &  0.1947   & 0.1852  &   0.1907  &  0.2076  & 7.918     &    3.919 & Cesetti et al. \\
NGC636    &  0.1850   & 0.1540  &   0.1416  &  0.2238  &7.559      &   4.034  & Cesetti et al. \\
NGC1425  &  -0.2152  &  0.0255 &   0.0720  &  -0.0077 &5.077	    &    2.385  & Cesetti et al. \\
NGC3115  &    0.3271  &  0.2211 &   0.1574	&   0.5500 & 8.550     &   5.756  & Cesetti et al. \\
NGC1700  &    0.2097 &   0.1699  &  0.2270	 &  0.2983 &  8.351    &    4.325  & Cesetti et al. \\
NGC1357   &   0.0744 &   0.2027  &  0.1718	 &  0.2265 & 6.914     &   3.572  & Cesetti et al. \\
NGC2613  &  -0.0003 &  0.1159  &   0.1126  &   0.5042 & 6.254     &   4.694 & Cesetti et al. \\
XSG1        &     -           &  0.40     &   0.20      &  0.70 &  7.841        &   5.589  &  \citet{LaBarbera16a} \\
XSG2        &     -           &  0.25     &   0.08      &  0.45 &  7.24        &   5.04  &  \citet{LaBarbera16a} \\

   \hline
\end{tabular}
%\tablefoot{
%\medskip
\end{table*}

\end{document}